\documentclass[12pt]{article}
\usepackage{amsmath,amssymb,bm,graphicx}
\usepackage[top=25truemm,bottom=25truemm,left=25truemm,right=25truemm]{geometry}
\allowdisplaybreaks

\title{Performance evaluation of matrix factorization for fMRI data}

\author{Yusuke Endo\thanks{Department of Mechanical Systems Engineering, Graduate School of Science and Engineering, Ibaraki University} 
\and Koujin Takeda\thanks{Department of Mechanical Systems Engineering, Graduate School of Science and Engineering, Ibaraki University, e-mail: koujin.takeda.kt@vc.ibaraki.ac.jp}}

\date{14th December, 2023}

\newcommand{\argmin}{\mathop{\rm arg~min}\limits}

\begin{document}
\maketitle

\begin{abstract}
In the study of the brain, there is a hypothesis that sparse coding is realized in information representation of external stimuli, which is experimentally confirmed for visual stimulus recently. However, unlike the specific functional region in the brain, sparse coding in information processing in the whole brain has not been clarified sufficiently. In this study, we investigate the validity of sparse coding in the whole human brain by applying various matrix factorization methods to functional magnetic resonance imaging data of neural activities in the whole human brain.
The result suggests sparse coding hypothesis in information representation in the whole human brain, because extracted features from sparse MF method, SparsePCA or MOD under high sparsity setting, or approximate sparse MF method, FastICA, can classify external visual stimuli more accurately than non-sparse MF method or sparse MF method 
under low sparsity setting.
\end{abstract}

Keywords: sparse coding, matrix factorization, fMRI, sparsity

%%%%%%%%%%%%%%%%%%%%%
\section{Introduction}
\label{sec1}

For information representation in the brain, various models are proposed based on the principle of reduction of information redundancy. 
In particular, it is widely recognized that sparse coding will be realized in the information representation of the visual 
and auditory stimuli \cite{Olshausen_sparsecoding, exA, exB, exC, exD, exE}. Recently, by physiological 
experiments with two-photon microscope imaging, it is found that visual stimulus can be reconstructed
from activities of a small number of neurons in the primary visual cortex \cite{Ohki}.
Overall, information representation in the specific functional region in the brain has been intensively investigated, 
whereas the representation in the whole brain has not been clarified sufficiently.

In this study, we attempt to model sparse coding in the whole human brain 
by applying matrix factorization (MF) to functional magnetic resonance imaging (fMRI) data, and characterize the 
feature in fMRI data by MF methods. 
There are several previous works on the analysis of fMRI data by MF methods.
First, independent component analysis (ICA) \cite{ICA} has been applied to fMRI data, 
where ICA is regarded as one of MF methods.
Among such studies by ICA, it is found that ICA can identify the resting state pattern in the brain \cite{restICA}
from resting-state fMRI data.
Next, there is also an application of MF method with sparse constraint (called sparse MF hereafter) to fMRI data \cite{fMRIDL}.
In this study, they pointed out that sparse MF can extract feature for a specific task of test subject 
from task-related fMRI data more appropriately than ICA, whereas ICA is better for extracting feature in 
resting state of the brain from resting-state fMRI data. It should be noted that they used synthetic fMRI data in their study.
As other previous work \cite{exD}, binary classification is performed 
for real experimental fMRI data to identify true action 
of test subject (for example, watching video or not) using extracted feature by
MF methods including sparse MF.

Based on these previous works, 
we attempt to validate sparse coding hypothesis through
classification of task-related fMRI data by applying various MF methods.
In our study, we apply {\it SparsePCA} \cite{SparsePCA} and {\it method of optimal directions} (MOD) \cite{MOD}
as sparse MF methods. We also conduct numerical MF analysis by 
principal component analysis (PCA) and ICA
(FastICA \cite{FastICA} in particular), 
which are conventional MF methods without sparse constraint.
However, while FastICA does not directly impose sparse constraint on the factorized matrix, it is known to maximize the kurtosis of the factorized matrix, resulting in approximate sparse factorized matrix.
 By comparing the results from various MF methods, 
 we aim to validate sparse coding for information processing in the whole human brain.
In our MF analysis, we attempt to classify true external visual stimulus in task-related fMRI data correctly 
using only extracted features by MF.  
In addition, to reinforce our argument from another point of view, 
we measure correlation between time series of extracted features by MF and 
time series of specific visual stimulus.
As expected, classification accuracy and correlation will depend on MF method.
For validation of our analysis, we also measure reconstruction error of MF.
Then we discuss which MF method is the most appropriate to model coding in fMRI data
through quantitative analysis of MF results. 

Finally, we should summarize the difference of our work from the previous studies. 
First, we mainly attempt to validate sparse coding hypothesis by applying various MF methods
including sparse ones,
while ICA analysis focuses on the signal source separation by independence.
Second, we use real public experimental data for classification of task by MF, whereas
the synthetic data was used in the previous work \cite{fMRIDL}.
Third, we focus on the visual stimulus and perform multi-label classification to predict 
which picture the test subject watches, whereas in the previous work \cite{exD}
they performed simpler binary classification task using the feature by MF. Binary classification 
may not be sufficient for validation of sparse coding hypothesis.
By these differences and the detailed statistical discussion,
 we believe that our result of the analysis can validate sparse cording hypothesis in real
fMRI data or real information processing in the brain more strongly than the previous studies.

It is also noted that the purpose of this study is to verify whether the representation of information 
in the brain is sparse or not by comparing sparse MF and non-sparse MF methods.
Therefore, the purpose is not to improve the classification accuracy or the value of
correlation with specific visual stimulus.

%%%%%%%%%%%%%%%%%%%%%
\section{Matrix factorization}
\label{sec2}

Throughout this paper, we describe matrix as a bold uppercase letter such as $\bm{A}$ and vector as a bold lowercase letter such as $\bm{a}$. Scalar is denoted as small italic, and
elements of vector/matrix are written as italic lowercase letter such as $[a_1,\hdots, a_n]$.

MF is a method to represent an original matrix
 as the product of two matrices. 
In MF, the original matrix $\bm{Y} \in \mathbb{R}^{M \times N}$ is decomposed to matrices $\bm{D} \in \mathbb{R}^{M \times K}$ and $\bm{X} \in \mathbb{R}^{K \times N}$. These matrices are related by $\bm{Y} = \bm{D} \bm{X} + \bm{E}$, where the matrix $\bm{E} \in \mathbb{R}^{M \times N}$ represents algorithmic error due to MF method.
 Here, we classify numerical MF algorithms as follows.
 In some MF methods like PCA or ICA, one first calculates factorized matrix $\bm{X}$ with orthogonal constraint, then constructs another factorized matrix $\bm{D}$ by $\bm{D}=\bm{Y} \bm{X}^{T}$.
In other methods like SparsePCA or MOD, one calculates
 two factorized matrices $\bm{D}$ and $\bm{X}$ alternately from the original matrix $\bm{Y}$.

In this work, we apply several MF methods to fMRI data matrix, which corresponds to $\bm Y$ in our notation.
The details of MF methods are elucidated in the following.

%%%%%%%%%%
\subsubsection*{PCA}
\label{sec2.1}

Widely-used PCA is regarded as an MF method
 to identify the orthogonal projection matrix  
 $\bm{X}^T \in \mathbb{R}^{N\times K}$, which is for removal of correlation between elements 
 in each column vector in the original matrix $\bm{Y}$.
Each column of the projection matrix $\bm{x}_i$ is obtained by solving the maximization problem in equation (\ref{PCA_1}) under the normalization constraint $\bm{x}_{i}^{T}\bm{x}_{i}=1$,
\begin{equation}
\label{PCA_1}
\hat{\bm{x}}_{i} = \underset{\bm{x}_i}{\rm argmax} \left( \frac{1}{M} \Sigma_{j=1}^{M} \{ \bm{y}_{j} \bm{x}_i^{T} - \bar{\bm{y}} \bm{x}_i^{T} \}^2 \right), 
\end{equation}
where $\bar{\bm{y}} = \frac{1}{M} \Sigma_{j=1}^{M} \bm{y}_j$ and $\bm y_j$ is the $j$-th column of $\bm Y$.
The optimal solution $\hat{\bm{x}}_i$ is obtained as eigenvector of covariance matrix of $\bm{Y}$.

%%%%%%%%%%
\subsubsection*{ICA and FastICA}
\label{sec2.2}

ICA was proposed as a signal separation method and has been applied to a wide range of fields such as speech recognition and image processing. Various methods of ICA have been proposed for different purposes, 
and the most well-known method, FastICA, is used in this study.
In FastICA, the factorized matrices $\bm D, \bm X$ are evaluated by the following procedure.

\begin{description}
\item[\small Pre-whitening]\mbox{}\\
The whitening matrix $\bm Q \in \mathbb{R}^{K \times M}$ is calculated by the following equation,
\begin{equation}
\bm Q = \bm \Sigma^{-1} \bm U^{T}.
\end{equation}
The matrices $\bm U, \bm \Sigma$ are computed by the singular value decomposition of $\bm Y$ as $\bm Y=\bm U  \bm \Sigma \bm V^{T}$.
The matrix $\bm U$ is the collection of left singular vectors of $\bm Y$
after shifting the mean value of the elements in $\bm y_i$ to zero for all $i$. The matrix $\bm \Sigma$ is the singular matrix of $\bm Y$.
After computing $\bm Q$, the original matrix $\bm Y$ is whitened to $\bm Z$ by multiplying the matrix $\bm Q$ as $\bm Z = \bm Q \bm Y$.

This whitening procedure is also called PCA whitening because it can be regarded as eigenvalue decomposition 
of the covariance matrix.

\item[\small Optimization]\mbox{}\\
Next, the orthogonal matrix $\bm W \in \mathbb{R}^{K \times K}$ must be obtained, which is also called unmixing matrix in the literature of ICA. The factorized matrix $\bm X$ is evaluated by multiplying $\bm W^{T}$ to the whitened matrix $\bm Z$ 
as $\bm X = \bm W^{T} \bm Z$.
This manipulation is for non-gaussianity maximization of column vectors in $\bm X$. 
In FastICA, the approximation of negentropy in equation (\ref{FastICA_1}) is used as the measure of non-gaussianity.
\begin{equation}
\label{FastICA_1}
\mathcal{J}_f (\bm{w}_i) = \left( \frac{1}{M} \Sigma_{j=1}^{M} G(\bm{w}_i^T \bm{z}_{j}) - \frac{1}{M} \Sigma_{j=1}^{M} G(\nu_j) \right)^2,
\end{equation}
where $\nu_j$ ($j \in \{1, \ldots, M\}$) is normal gaussian random variable. $G(.)$ is a non-linear function in general, and $G(x)=-\exp(-x^2 /2)$ is used in this study. To maximize $\mathcal{J}_f$ with respect to $\bm{w}_i$, 
the iterative algorithm for fixed-point is used, which is expressed as update rule of $\bm{w}_i$,
\begin{equation}
\label{FastICA_2}
\bm{w}_i \leftarrow \frac{1}{M} \Sigma_{j=1}^M \bm{z}_j G'(\bm{w}_i^T \bm{z}_{j}) - G''(\bm{w}_i^T \bm{z}_{j}) \bm{w}_i.
\end{equation}
$G'(.)$ and $G''(.)$ represent the first- and the second-order derivatives of $G(.)$, respectively. After convergence,
the vectors $\bm{w}_i$'s are orthogonalized and their norms are normalized to 1.

\item[\small Computation of factorized matrices]\mbox{}\\
After pre-whitening and optimization, the factorized matrices $\bm D, \bm X$ are finally evaluated as follows,
\begin{eqnarray}
\bm X &=& \bm W^{T} \bm Q \bm Y, \nonumber \\
\bm D &=& \bm Q^{+} \bm W,
\end{eqnarray}
where $\bm Q^{+}$ is the pseudo-inverse of the matrix $\bm Q$.

\end{description} 

%%%%%%%%%%
\subsubsection*{SparsePCA}
\label{sec2.3}

In SparsePCA \cite{SparsePCA}, the projection matrix $\bm X^{T}$ is constructed sparsely by adding a regularization term to the cost function of PCA. Due to sparsity, the result of SparsePCA can be interpreted much easily than the original PCA. Here, elastic net is used as the regularization method in SparsePCA. Accordingly, the cost function of SparsePCA is given as
\begin{equation}
\label{SparsePCA_1}
\mathcal{J}_s (\bm{V}, \bm{W}) = \Sigma_{j=1}^N ||\bm{y}_j - \bm{y}_j \bm{W}^T \bm{V} ||_{2}^2 + \lambda \Sigma_{i=1}^{K} ||\bm{w}_i||_1 + \gamma \Sigma_{i=1}^{K} ||\bm{w}_i||_{2}^{2}.
\end{equation}
The parameters $\lambda$ and $\gamma$ are for controlling sparsity. 
The matrices $\bm{V},\bm{W}\in\mathbb{R}^{K \times N}$ are for calculation of the projection matrix $\bm{X}^T$. 
The cost function $\mathcal{J}_s$ is minimized under the constraint $\bm{V}^T\bm{V}=\bm{I}_N$, where $\bm I_N$ is $N$-dimensional identity matrix. 
Roughly speaking, in SparsePCA either matrix $\bm{V}$ or $\bm{W}$ is fixed first and another matrix is calculated as the minimizer of $\mathcal{J}_s$, then the roles of two matrices $\bm V, \bm W$ are exchanged and the minimization of $\mathcal{J}_s$ is repeated.
More precisely, at the $n$th step the minimization problem with respect to $\bm{W}$ is solved under fixed $\bm{V}$, which is computed at the previous $(n-1)$th step.
 The minimization for $\bm{W}$ is solved by the methods such as LARS \cite{LARS}, because it can be reformulated as LASSO 
 regression \cite{LASSO}. After computing $\bm{W}$, the matrix $\bm{V}$ minimizing $\mathcal{J}_s$ is calculated by fixing $\bm{W}$. The optimal solution for $\bm{V}$ is given by singular value decomposition of $\bm{W} \bm{Y}^T \bm{Y}$. After convergence of alternate estimation for $\bm V, \bm W$, the projection matrix $\bm X^{T}$ is obtained from normalized 
$\bm W$. Another factorized matrix $\bm D$ is obtained in the same manner as PCA.

%%%%%%%%%%
\subsubsection*{MOD}
\label{sec2.4}

MOD \cite{MOD} is one of dictionary learning methods, where the dictionary matrix $\bm{D}$ and the sparse representation matrix $\bm{X}$ are evaluated alternately. In this method, the matrices $\bm{D}$ and $\bm{X}$ are computed by solving the minimization problem as below.
\begin{equation}
\label{MOD_1}
\big\{ \bm{D}, \{ \bm{x}_j \}_{j=1}^N \big\} = \argmin_{\bm{D}, \{ \bm{x}_j\}_{j=1}^{N}} ||\bm{y}_j - \bm{D} \bm{x}_j||_{2}^{2} \quad s.t. \quad ||\bm{x}_j||_0 \leq k_0,\ 1 \leq j \leq N,
\end{equation}
where $||.||_0$ is $l_0$-norm and represents the number of non-zero elements. 
By tuning the parameter $k_0$ for sparsity, one can control maximum number of non-zero elements 
in sparse factorized matrix $\bm X$.
For the solution, the
 minimization with respect to $\bm{D}$ or $\bm{X}$ is conducted alternately like SparsePCA.
For the calculation of $\bm{X}$, orthogonal matching pursuit \cite{OMP1,OMP2} is used.
The dictionary $\bm{D}$ is computed by Moore-Penrose inverse as 
$\bm{Y}\bm{X}^T\{\bm{XX}^T\}^{-1}$.

%%%%%%%%%%%%%%%%%%%%%
\section{Numerical analysis}
\label{sec3}

%%%%%%%%%% 
\subsection{fMRI dataset}
\label{sec3.1}

\begin{figure}[t]
	 \begin{center}
      \includegraphics[scale=0.58]{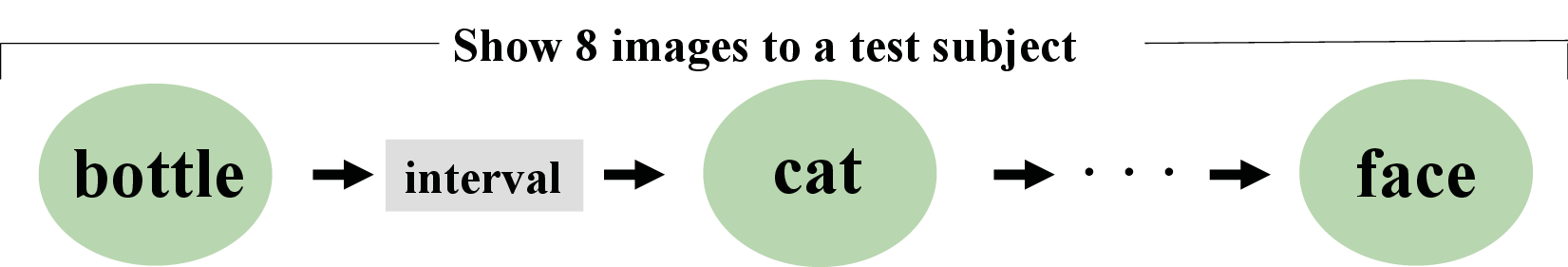}
	  \caption{Data acquisition of Haxby dataset}
	  \label{Haxby_experiment}
	 \end{center}
\end{figure}

In our numerical analysis, we use Haxby dataset \cite{Haxby_data}, which is an fMRI dataset observing human response to image.
 In data acquisition of Haxby dataset,
 one of 8 images is shown to a test subject for a while, then the next image is shown after an interval and exchange of images (Figure \ref{Haxby_experiment}). The 8 images are as follows: shoe, house, scissors, scrambledpix, face, bottle, cat, and chair.
In one trial of data acquisition, all 8 images are shown sequentially to a test subject. The data of 12 trials for 6 test subjects are included in the whole Haxby dataset.
In our MF analysis, we use fMRI data from 3 test subjects, Nos. 2, 3, and 4 in this dataset.
 For each test subject we analyze data of one trial, which is 
acquired from 39964 voxels within $121$ time steps. Namely, the size of the fMRI data matrix is $121\times 39964$, or $\bm{Y}\in\mathbb{R}^{121\times 39964}$.

\begin{figure}[t]
	 \begin{center}
      \includegraphics[scale=0.7]{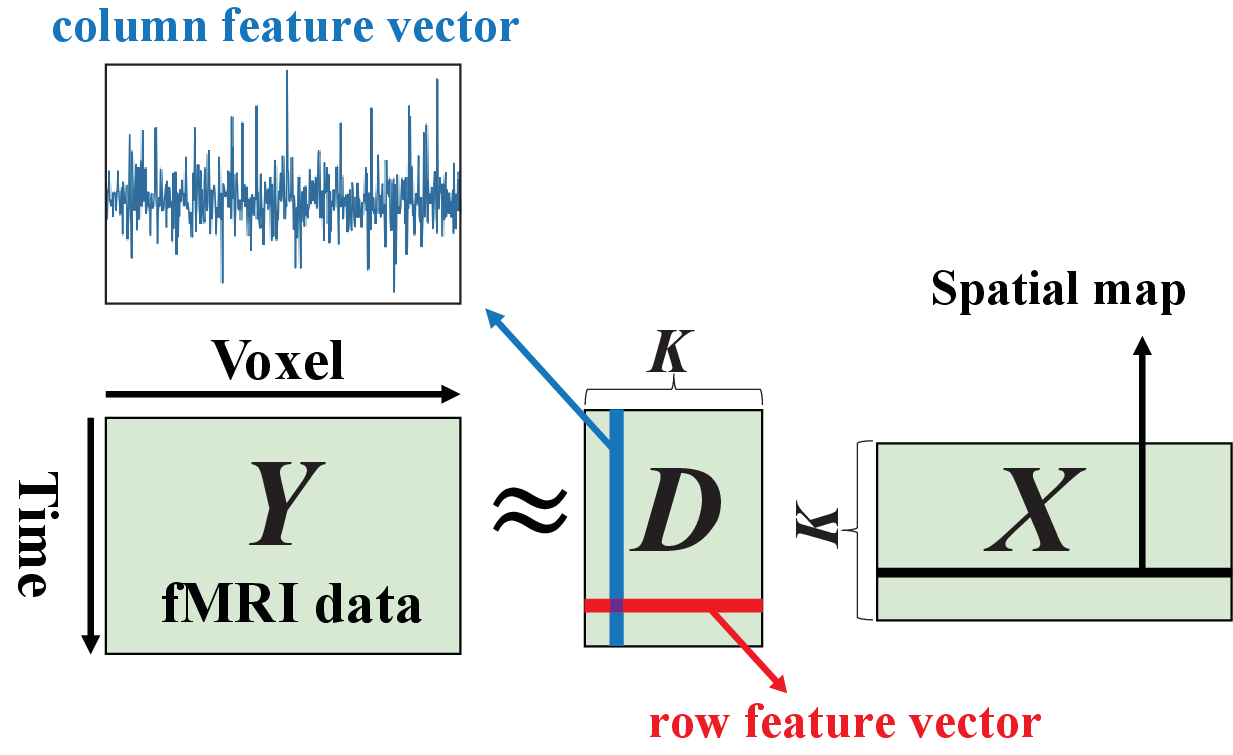}
	  \caption{Scheme of MF for fMRI data}
	  \label{MF_fmri}
	 \end{center}
\end{figure}

%%%%%%%%%%
\subsection{Performance evaluation of matrix factorization}
\label{sec3.2}

The procedure of MF in our study is sketched in Figure \ref{MF_fmri}. Each row in factorized matrix $\bm{D}$ is regarded as an extracted feature vector at a specific time step, which is used as an input for classifier.
 Each column in $\bm{D}$ is a temporal feature vector, which will
also give the information of the external stimulus in the experiment and is used for evaluation of correlation. 
 For performance evaluation of MF, classification accuracy, correlation between temporal vector in $\bm D$ and 
 the external visual stimulus, and reconstruction error of MF are used as explained below.

%%%%%%%%%%
\subsubsection*{Classification accuracy}
\label{sec3.2.1}

\begin{figure}[t]
	 \begin{center}
      \includegraphics[scale=0.6]{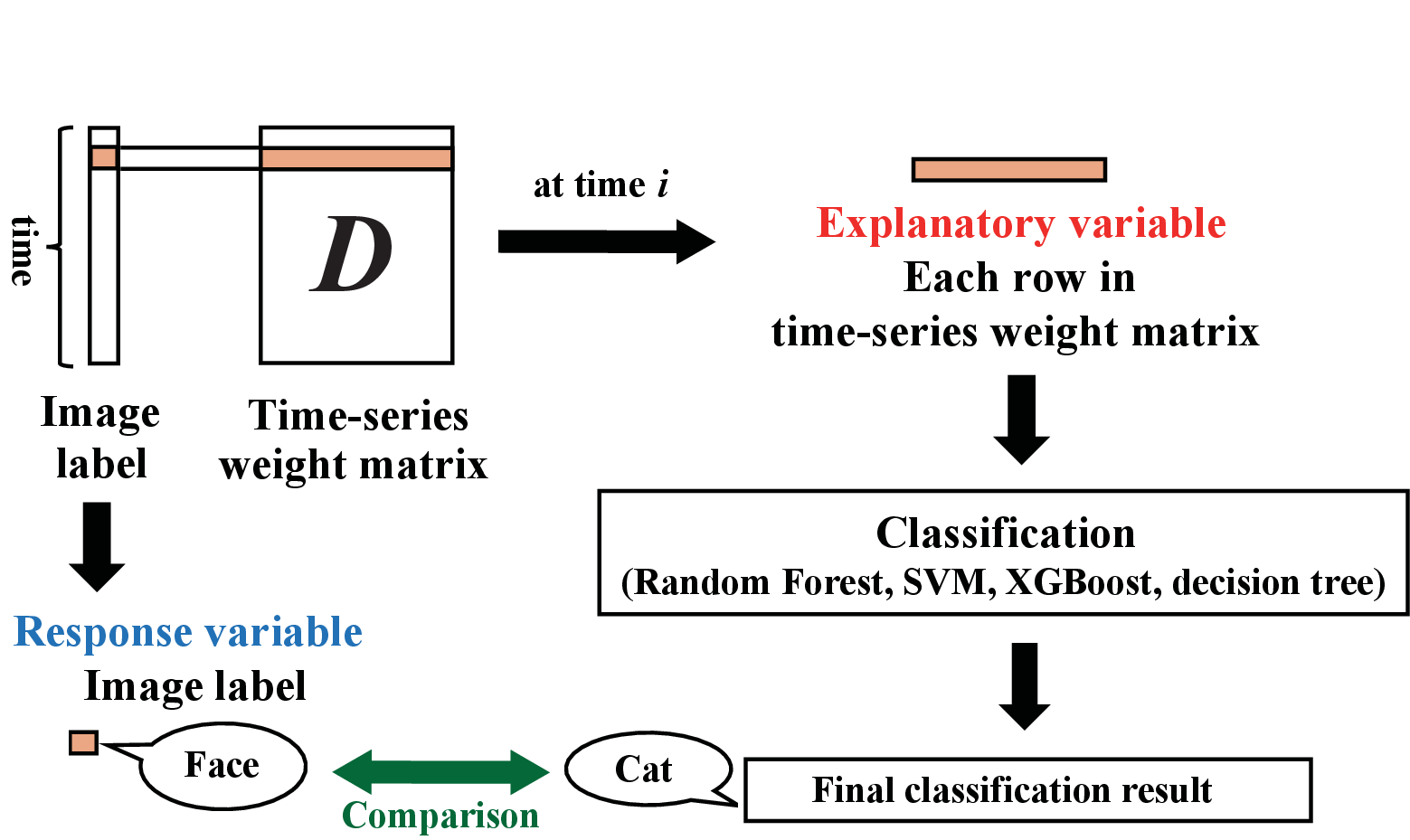}
	  \caption{Scheme of classification}
	  \label{class_MF}
	 \end{center}
\end{figure}
We attempt to classify visual stimuli using factorized matrix $\bm{D}$ to reveal which MF method can 
appropriately extract features of true visual stimulus. 

The scheme of the classification is sketched in Figure \ref{class_MF}. 
For classification, row vector at a specific time step in $\bm{D}$ is regarded as explanatory variable
 and label of image at a specific time step as response variable in statistical meaning. 
 After extracting row vector at a specific time, we put it into classifiers.
In our analysis we use four classifiers: random forest, support vector machine (SVM) \cite{SVM}, XGBoost \cite{XGBoost}, and decision tree.
To implement these classifiers, we use scikit-learn library (version 1.2.2) for Python (version 3.10.11).
We vary some parameters or settings in these classifiers to find better classification result for each classifier,
which is mentioned in the part of classification result (section \ref{sec4.1}). Other parameters are set as default values. 
For evaluation of classification accuracy, we randomly shuffled the row vectors for 121 time steps in $\bm Y$, then
 we perform 5-fold cross validation by dividing fMRI data matrix $\bm Y$ into 5 sections in time axis.

Additionally, in our setting, spatial feature matrix $\bm X$ is assumed to be sparse under sparse MF methods.
To verify the significance of spatial sparsity in fMRI data,
we also factorize the {\it transpose} matrix $\bm Y^T$ and perform MF analysis.
 In this case, the roles of row/column vectors are exchanged:
 dense matrix $\bm D$ and sparse matrix $\bm X$ correspond to spatial and temporal features, respectively.

%%%%%%%%%%
\subsubsection*{Correlation}
\label{sec3.2.2}

As explained, there exists a time period when the subject is shown a particular image.
Then we expect that a specific temporal feature vector in a column in $\bm D$ may represent the feature of the visual stimulus from a particular image.
To verify whether each MF method can extract feature of visual stimulus appropriately or not, we 
evaluate the correlation between a specific column vector $\bm d$ in the factorized matrix $\bm{D}$
 and temporal vector of stimulus from the true image. The definition of correlation for a given image label is given as follows.
\begin{equation}
\label{eq:correlation}
{\rm correlation}\ (\bm d, \bm s^{\rm IMG}) = 
\frac{  \left| (\bm d - \overline{ \bm d})^T (\bm s^{\rm IMG} - \overline{\bm s^{\rm IMG}})  \right| }
{  \left\| \bm d - \overline{\bm d} \right\|_2  \left\| (\bm s^{\rm IMG} - \overline{\bm s^{\rm IMG}}) \right\|_2 }.
\end{equation}
In this definition, $\| \cdot \|_2$ is $\ell_2$-norm and overline means the arithmetic average of all elements in the vector. 
The vector $\bm s^{\rm IMG}$ 
is the column vector representing stimulus for a given image denoted by IMG $\in\{$shoe, house, scissors, scrambledpix, face, bottle, cat, chair, rest$\}$, where rest means that no image is shown to test subject. The element of $\bm s^{\rm IMG}$ at time step $i$ is
defined as 
\begin{eqnarray}
s_i^{\rm IMG} &=& \mathbb{I} ( i \in [t_{\rm start}^{\rm IMG}- \delta t, t_{\rm end}^{\rm IMG}+ \delta t ] ) \ \ \  {\rm for} \ \ {\rm IMG} \neq {\rm rest}, \\
 s_i^{\rm rest} &=&
 \begin{cases}
  1 & {\rm if} \  s_i^{\rm IMG'} = 0 \ \ \ {\rm for\ all \ \ IMG' \ \ excepting \  rest},  \\
  0 & {\rm otherwise},
 \end{cases}
\end{eqnarray}
where
$\mathbb{I} [A]$ is indicator function defined by $\mathbb{I} [A] = 1$ for true proposition $A$ and $0$ for false, 
and $t_{\rm start}^{\rm IMG}, t_{\rm end}^{\rm IMG}$ 
are the time to start/finish showing a particular image IMG to the subject, respectively.  
The time interval $\delta t$ is for making stimulus input time period slightly longer for convenience of analysis, and we set $\delta t=4$ in our analysis.
 
The value of correlation means the synchronization between a specific temporal feature vector $\bm d$ 
in $\bm D$ and a visual stimulus by a particular image. 
When the correlation is larger, it means that the extraction of feature vector 
by MF works more appropriately than the case of smaller correlation.

%%%%%%%%%% 
\subsubsection*{Reconstruction error}
\label{sec3.2.3}

We also investigate which MF method can keep the information of the original matrix $\bm Y$ %after factorization 
by observing algorithmic error $\bm E$. The reconstruction error matrix $\bm \Delta$ is defined by taking absolute value of each element in $\bm E$, namely $\delta_{ij} = | e_{ij} |$ or as in equation (\ref{reconstruction error}),
\begin{equation}
\label{reconstruction error}
\bm \Delta =
\left[
\begin{matrix}
|y_{11}-\sum_{i=1}^{K} d_{1i} x_{i1}| & |y_{12} - \sum_{i=1}^{K} d_{1i} x_{i2}| & \hdots & |y_{1N} - \sum_{i=1}^{K} d_{1i} x_{iN}|\\
|y_{21}-\sum_{i=1}^{K} d_{2i} x_{i1}| & |y_{22} - \sum_{i=1}^{K} d_{2i} x_{i2}| & \hdots & |y_{2N} - \sum_{i=1}^{K} d_{2i} x_{iN}|\\
\vdots & \vdots & \ddots & \vdots\\
|y_{M1}-\sum_{i=1}^{K} d_{Mi} x_{i1}| & |y_{M2} - \sum_{i=1}^{K} d_{Mi} x_{i2}| & \hdots & |y_{MN} - \sum_{i=1}^{K} d_{Mi} x_{iN}|
\end{matrix}
\right].
\end{equation}

In this study, statistics of $\delta_{ij}$ (= elements in $\bm \Delta$) and Frobenius norm of $\bm \Delta$ are used as the indices of the reconstruction error. The statistical indices of reconstruction error such as maximum, median, and mean are desired to be small. The extracted feature will be inappropriate when they are too large.
For evaluation of reconstruction error, we do not take any ensemble averages in our analysis.

%%%%%%%%%%%%%%%%%%%%%
\section{Result}
\label{sec4}

%%%%%%%%%%
\subsection{Classification accuracy}
\label{sec4.1}

\begin{figure}[htbp]
	 \begin{center}
      \includegraphics[scale=0.55]{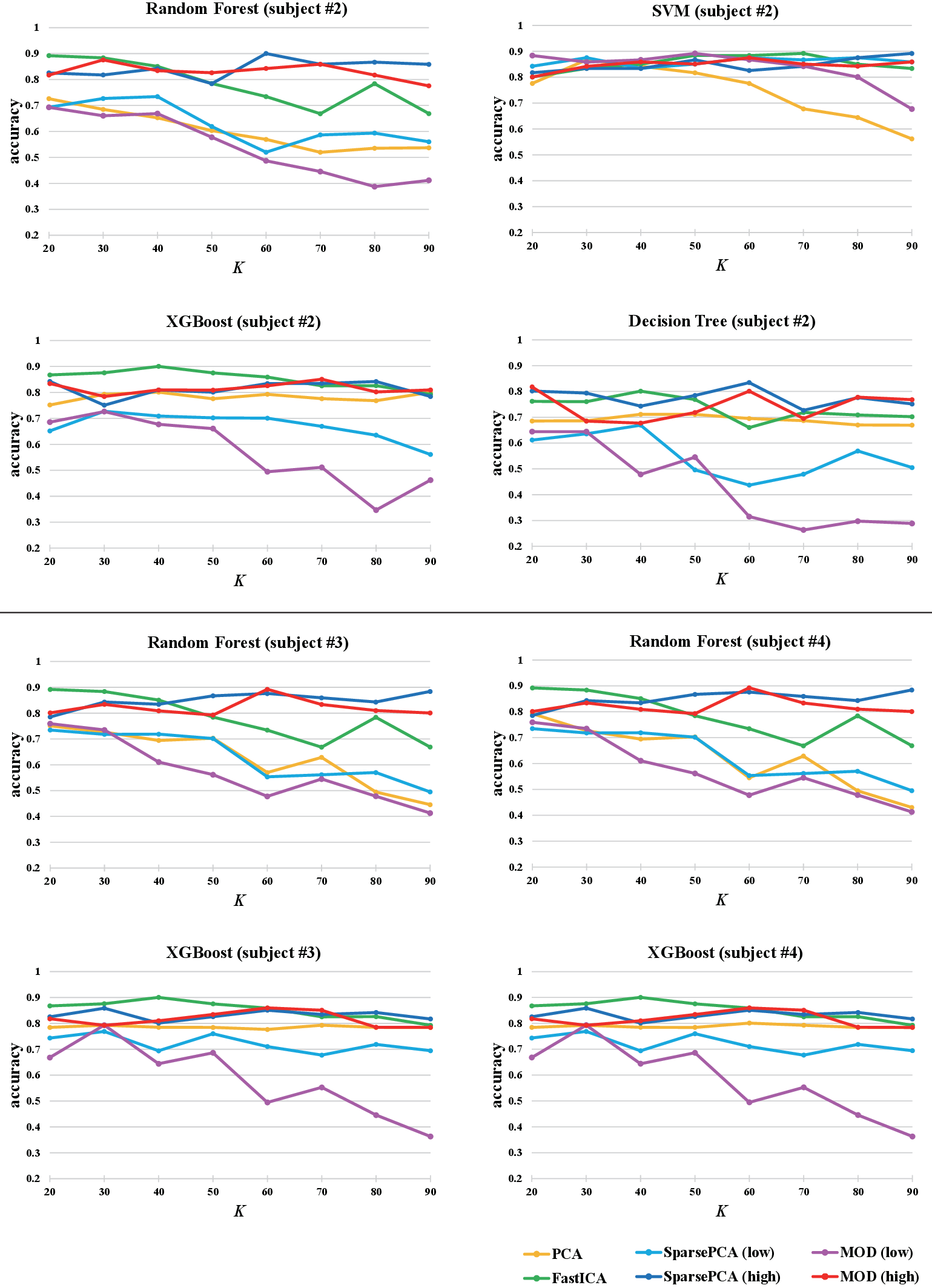}
	  \caption{Classification accuracy versus $K$: The results by PCA
	  , FastICA, SparsePCA under $\lambda=0.1$ (low) and $1$ (high), and MOD 
	  under $k_0/K=0.1$ (high) and $0.75$ (low) are depicted.
	  The results by four classifiers, random forest and XGBoost with 10 trees and 7 depth, SVM with linear kernel,
	  and decision tree with 10 depth are shown for the data of test subject No. 2. 
	  The results for the data of three test subjects Nos. 2, 3, and 4 are also compared under random forest 
	  and XGBoost with 10 trees and 7 depth.}
	  \label{classification_result}
	 \end{center}
\end{figure}

First, we show the classification result by changing MF method. 
We perform the analysis for the data from test subject No. 2 by four classifiers: 
random forest (with 10 trees and 7 depth), 
SVM (with linear kernel), XGBoost (with 10 trees and 7 depth), and decision tree (with 10 depth).
The result of classification with varying $K$ (= the parameter of matrix size) is depicted
in Figure \ref{classification_result}.
 Sparse parameter $\lambda$ is set to 0.1 (low sparsity setting, denoted by low) or 1 (high sparsity setting, denoted by high) for SparsePCA.  Another parameter $\gamma$ in SparsePCA 
 is set to 0.01.
 Fraction of zero elements in factorized matrix, namely $k_0/K$, is set to 0.1 (high sparsity setting, denoted by high) or 0.75 (low sparsity setting, denoted by low) for MOD.

Summary of the result is given in the following.
 Under random forest classifier, classification accuracy tends to decrease with increasing $K$ in general. 
 Exceptionally, sparse MF methods (SparsePCA, MOD) 
 with high sparsity setting can keep high accuracy even with large $K$,
  whereas other methods do not.
 For SVM, in addition to sparse MF methods with high sparsity setting, FastICA and SparsePCA with low sparsity setting 
 also give high accuracy. 
 Under XGBoost and decision tree, only sparse MF methods with low sparsity setting yield lower accuracy.
 From these observations, it can be concluded that sparse MF methods  
 with high sparsity setting and FastICA give higher accuracy than other methods, 
 and this result is independent of the classifier. 
 In particular, for larger matrix dimension $K$, sparse MF methods often
 give higher accuracy than FastICA. This result suggests that sparsity is very significant in feature extraction 
 from fMRI data, in addition to the statistical independence by FastICA suggested in the previous studies.
 The dependence of these methods on the dimension $K$ will be discussed again in the last part of this subsection. 
 
 The advantage of sparse MF methods and FastICA may be explained as follows.  
 In our setting by sparse MF methods, sparsity is directly imposed on the spatial matrix $\bm X$.
 In contrast, sparsity is not directly imposed in FastICA and statistical independence is imposed instead. 
  However, it is found that many matrix elements take values near zero and
   approximate sparse matrix is realized eventually by FastICA, 
   which is statistically verified in the subsequent subsection.
  Therefore, higher accuracy by these methods suggests that the signals 
 by the visual stimulus are localized in several regions in the whole brain. In other words,
 visual stimulus may be expressed sparsely in the information processing in the brain.
 
 Next, we move on to the details of the dependence on the classifier. 
 First, the reason for the high accuracy by PCA, which is not a sparse MF method, only under XGBoost 
 and decision tree is discussed.
 Significant features in PCA are mainly included in eigenvectors of several larger eigenvalues.  
 In XGBoost and decision tree, the importances of all feature vectors are evaluated at first
 for constructing trees. Hence, the feature vectors with larger eigenvalues in PCA are mainly used in these classifiers.
 Such construction of trees helps to keep high accuracy even under large $K$.
 In contrast, in random forest, trees are constructed using features extracted by bootstrapping,
 where only randomly sampled features are used.
 Therefore, in random forest it may happen that important features with larger eigenvalues in PCA 
 are discarded under large $K$, which may degrade the accuracy.
 In SVM, large $K$ will make the feature space for classification of features by PCA more complex, 
 which is the cause of low accuracy.
 Even with SVM, it is expected that simpler feature space is realized under sparse MF method
 with high sparsity setting or FastICA, which may help to keep high accuracy by these methods under large $K$.
 
We also perform the analysis for the data from different test subjects
to verify that the result does not depend on the individual subject.
 We use the data from 3 test subjects in Haxby dataset (subjects Nos. 2, 3, and 4), where
random forest and XGBoost with 10 trees and 7 depth is used as classifier.
As shown in Figure \ref{classification_result}, no significant difference among subjects is observed
in all MF methods. 

\begin{figure}[htbp]
	 \begin{center}
      \includegraphics[scale=0.75]{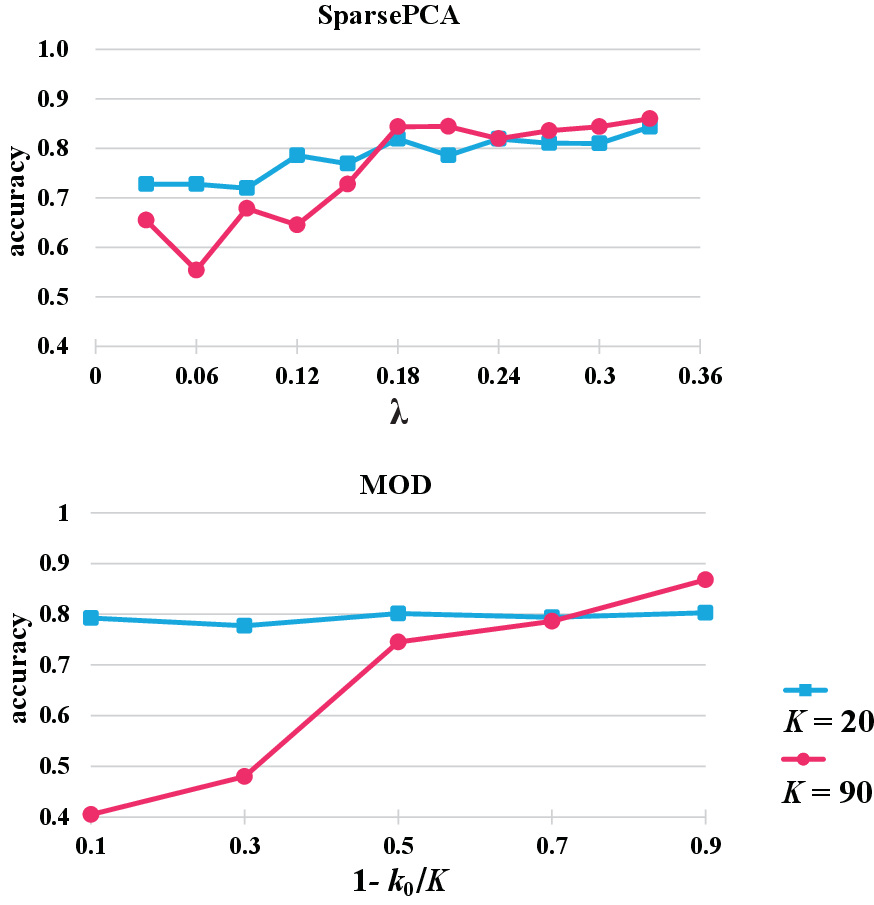}
	  \caption{Classification accuracies by SparsePCA and MOD under $K=20, 90$: Parameter $\lambda$ in SparsePCA is varied
	  for controlling sparsity in sparse factorized matrix. The fraction of zero elements in factorized matrix 
	  is changed by tuning parameter $k_0$ in MOD.}
	  \label{class_parameter}
	 \end{center}
	 \vspace{5mm}
	 \begin{center}
      \includegraphics[scale=0.75]{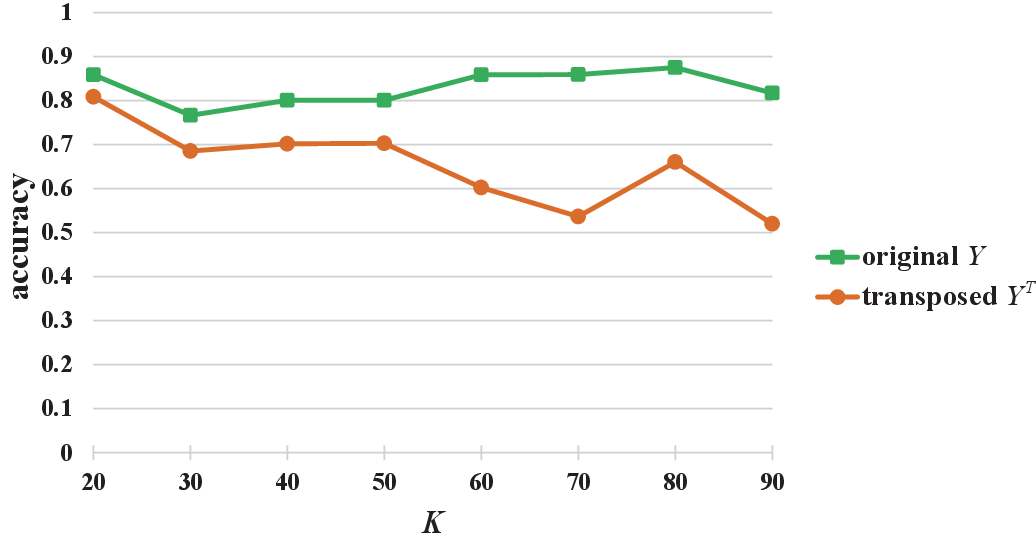}
	  \caption{Classification accuracies by SparsePCA for the original fMRI data $\bm Y$ and the transposed data $\bm Y^T$}
	  \label{class_spacetime}
	 \end{center}
\end{figure}

\begin{table}[htbp]
	 \centering
	 \caption{Dependence of accuracy on MF method and classifier $(K=20)$}
	 \label{tab:class1}
	 \scalebox{0.86}[0.86]{
	 \begin{tabular}{|c|c|cccccc|}
	 \hline
classifier & condition & PCA & FastICA & SparsePCA & SparsePCA & MOD & MOD \\
 & of classifier &  & & (low) & (high) & (low) & (high) \\ \hline \hline
random forest & \# of trees=10, depth=3	& 0.586 & 0.636 & 0.512 & 0.685 & 0.636 & 0.725 \\ \hline
random forest & \# of trees=10, depth=5	& 0.735 & 0.876 & 0.709 & 0.842 & 0.793 & 0.851 \\ \hline
random forest & \# of trees=10, depth=7	& 0.751 & 0.892 & 0.726 & 0.809 & 0.802 & 0.842 \\ \hline
random forest & \# of trees=50, depth=3	& 0.520 & 0.751 & 0.552 & 0.750 & 0.676 & 0.775 \\ \hline
random forest & \# of trees=100, depth=3 & 0.545 & 0.693 & 0.519 & 0.784 & 0.685 & 0.775 \\ \hline \hline 
SVM & RBF kernel & 0.685 & 0.726 & 0.834 & 0.635 & 0.734 & 0.610 \\ \hline
SVM & linear kernel & 0.776 & 0.801 & 0.842 & 0.817 & 0.884 & 0.801 \\ \hline
SVM & polynomial kernel	& 0.438 & 0.726 & 0.603 & 0.603 & 0.693 & 0.710 \\ \hline \hline
XGBoost	& \# of trees=10, depth=3 & 0.875 & 0.867 & 0.759 & 0.859 & 0.826 & 0.875 \\ \hline
XGBoost	& \# of trees=10, depth=5 & 0.867 & 0.867 & 0.751 & 0.859 & 0.826 & 0.859 \\ \hline
XGBoost	& \# of trees=10, depth=7 & 0.867 & 0.867 & 0.751 & 0.859 & 0.826 & 0.859 \\ \hline
XGBoost	& \# of trees=50, depth=3 & 0.875 & 0.892 & 0.759 & 0.859 & 0.826 & 0.875 \\ \hline
XGBoost	& \# of trees=100, depth=3 & 0.875 & 0.892 & 0.759 & 0.859 & 0.826 & 0.875 \\ \hline \hline
decision tree & depth=5	& 0.594 & 0.563 & 0.529 & 0.594 & 0.611 & 0.710 \\ \hline
decision tree & depth=7	& 0.661 & 0.762 & 0.612 & 0.785 & 0.628 & 0.826 \\ \hline
decision tree & depth=10 & 0.685 & 0.762 & 0.612 & 0.802 & 0.644 & 0.818 \\ \hline
	 \end{tabular}
	 }
	 \vspace{5mm}
	 \centering
	 \caption{Dependence of accuracy on MF method and classifier $(K=90)$}
	 \label{tab:class2}
	 \scalebox{0.86}[0.86]{
	 \begin{tabular}{|c|c|cccccc|}
	 \hline
classifier & condition & PCA & FastICA & SparsePCA & SparsePCA & MOD & MOD \\
& of classifier & & & (low) & (high) & (low) & (high) \\ \hline \hline
random forest & \# of trees=10, depth=3	& 0.437 & 0.578 & 0.405 & 0.677 & 0.405 & 0.701 \\ \hline
random forest & \# of trees=10, depth=5	& 0.495 & 0.676 & 0.561 & 0.859 & 0.454 & 0.851 \\ \hline
random forest & \# of trees=10, depth=7	& 0.529 & 0.669 & 0.553 & 0.809 & 0.396 & 0.884 \\ \hline
random forest & \# of trees=50, depth=3	& 0.404 & 0.536 & 0.421 & 0.726 & 0.404 & 0.792 \\ \hline
random forest & \# of trees=100, depth=3 & 0.404 & 0.552 & 0.429 & 0.784 & 0.404 & 0.784 \\ \hline \hline
SVM & RBF kernel & 0.404 & 0.479 & 0.553 & 0.750 & 0.421 & 0.750 \\ \hline
SVM & linear kernel & 0.562 & 0.834 & 0.859 & 0.892 & 0.677 & 0.908 \\ \hline 
SVM & polynomial kernel	& 0.404 & 0.421 & 0.413 & 0.635 & 0.404 & 0.826 \\ \hline \hline
XGBoost & \# of trees=10, depth=3 & 0.850 & 0.793 & 0.677 & 0.809 & 0.708 & 0.876 \\ \hline
XGBoost & \# of trees=10, depth=5 & 0.842 & 0.793 & 0.685 & 0.826 & 0.627 & 0.876 \\ \hline
XGBoost & \# of trees=10, depth=7 & 0.842 & 0.793 & 0.677 & 0.826 & 0.618 & 0.876 \\ \hline
XGBoost	& \# of trees=10, depth=7 & 0.850 & 0.834 & 0.677 & 0.809 & 0.708 & 0.876 \\ \hline
XGBoost	& \# of trees=50, depth=3 & 0.850 & 0.842 & 0.677 & 0.809 & 0.708 & 0.876 \\ \hline \hline
decision tree & depth=5	& 0.595 & 0.570 & 0.505 & 0.520 & 0.256 & 0.570 \\ \hline
decision tree & depth=7	& 0.669 & 0.702 & 0.505 & 0.652 & 0.273 & 0.694 \\ \hline
decision tree & depth=10 & 0.669 & 0.702 & 0.505 & 0.751 & 0.232 & 0.735 \\ \hline \hline
	 \end{tabular}
	 }
\end{table}

In Figure \ref{class_parameter}, the dependence of classification accuracy on sparse parameter $\lambda$ in SparsePCA 
is shown for the data of test subject No. 2 under $K=20, 90, \gamma=0.01$, and random forest classifier with 10 trees and 7 depth. 
For smaller matrix size ($K=20$) the variation range of classification accuracy is at most 0.1, whereas for larger size ($K=90$) 
it increases to 0.25. 
In addition, the dependence of classification accuracy on the number of zero elements in MOD, namely $1 - k_0/K$, is shown under $K=20, 90$. 
Like SparsePCA, the variation range of classification accuracy is at most 0.1 for $K=20$, whereas it increases 
to 0.4 for $K=90$. These results indicate that sparse parameter $\lambda$ in SparsePCA or sparsity in MOD affects classification accuracy more significantly for larger $K$.
Equivalently, this means the robustness of classification accuracy
against the change of sparsity parameter in sparse MF methods when $K$ is small.

In addition, we also use the transposed matrix $\bm Y^T$ in classification, where
the roles of spatial and temporal features are exchanged. We perform MF analysis
by SparsePCA under $\lambda=1, \gamma=0.01$, and random forest classifier with 10 trees and 7 depth.
Obviously from Figure \ref{class_spacetime}, the transposed matrix yields lower accuracy, which indicates that
the spatial sparsity is significant in extracting feature from fMRI data.

Finally, we discuss the dependence on parameter in classifiers, which is summarized in Tables \ref{tab:class1}, \ref{tab:class2}. 
For tree-based methods (random forest, XGBoost, and decision tree), accuracy tends to improve by increasing the number or depth of trees. Under large number of or deeper trees, sparse MF method with high sparsity setting gives higher accuracy than others.
For SVM, linear kernel gives better accuracy than other kernels, radial based function (RBF) kernel (gaussian kernel in other words) and polynomial kernel, 
which implies that the feature space by MF is linearly separable.

In these tables, we also compare the results by sparse MF methods (SparsePCA, MOD) with high sparsity setting and FastICA. 
For $K=20$ these methods give almost the same accuracy, while for $K=90$ sparse MF methods often give
higher accuracy than FastICA. It should be noted that this property does not depend on the choice of classifier.
The reason will be explained as follows.
For smaller matrix dimension $K$, it is generally difficult to obtain sparse factorized matrix by MF due to 
the smaller number of factorized matrix elements, which results in almost the same classification accuracy by sparse MF method or FastICA. 
When we increase the dimension $K$, sparse factorized matrix can be obtained more easily due to the increase of the number of factorized matrix elements, which will enable to extract more appropriate feature by sparse MF method. This fact also suggests that sparsity is
significant in extracting appropriate features by MF from fMRI data. 

%%%%%%%%%%
\subsection{Correlation}
\label{sec4.2}

We evaluate the correlation between column feature vector in factorized matrix and temporal vector of a particular visual stimulus,
which is defined in equation (\ref{eq:correlation}). Here we use the data from subject No. 2.

In Tables \ref{tab:corr1}, \ref{tab:corr2}, and \ref{tab:corr3},
the value of maximum correlation among all feature vectors 
obtained by each MF method is shown for $K=20, 50,$ and $90$, respectively.  
From these tables, sparse MF methods (SparsePCA, MOD) with high sparsity setting 
tend to give the largest value of correlation under $K=50, 90$. For $K=20$, 
FastICA gives the largest value of correlation in almost all cases. However,
sparse MF methods also give large value, which is comparable to the result by FastICA.
 This implies that the information in the brain 
may be represented in low dimensional feature space (like $K=20$) or sparsely in high dimensional feature space
(like $K=50, 90$). Furthermore, the advantage of sparse MF methods under larger $K$ 
 is consistent with the result of accuracy in the last subsection.

\begin{table}[htbp]
	 \centering
	 \caption{Dependence of correlation on MF method and visual stimulus $(K=20)$}
	 \label{tab:corr1}
	 \scalebox{1.0}[1.0]{
	 \begin{tabular}{|c|cccccc|}
	 \hline
IMG & PCA & FastICA & SparsePCA & SparsePCA & MOD & MOD \\
&     & & (low) & (high) & (low) & (high) \\ \hline \hline
rest (no image) & 0.679 & \bf 0.881 & 0.539 & 0.820 & 0.600 & 0.716 \\ \hline
shoe & 0.883 & \bf 0.930 & 0.858 & 0.919 & 0.924 & 0.920 \\ \hline
house & 0.904 & \bf 0.934 & 0.807 & 0.923 & 0.921 & 0.931 \\ \hline
scissors & 0.869 & \bf 0.911 & 0.830 & 0.893 & 0.870 & 0.890 \\ \hline
scrambledpix & 0.637 & \bf 0.861 & 0.658 & 0.834 & 0.706 & 0.858 \\ \hline
face & 0.703 & 0.811 & 0.842 & 0.794 & 0.816 & \bf 0.933 \\ \hline
bottle & 0.921 & \bf 0.966 & 0.778 & 0.952 & 0.931 & 0.955 \\ \hline
cat & 0.885 & \bf 0.923 & 0.922 & 0.917 & 0.845 & 0.919 \\ \hline
chair & 0.897 & \bf 0.921 & 0.871 & 0.904 & 0.816 & 0.894 \\ \hline
	 \end{tabular}
	 }
	 \vspace{7mm}
	 \centering
	 \caption{Dependence of correlation on MF method and visual stimulus $(K=50)$}
	 \label{tab:corr2}
	 \scalebox{1.0}[1.0]{
	 \begin{tabular}{|c|cccccc|}
	 \hline
IMG & PCA & FastICA & SparsePCA & SparsePCA & MOD & MOD \\
&     & & (low) & (high) & (low) & (high) \\ \hline \hline
rest (no image)	& 0.679 & \bf 0.849 & 0.566 & 0.790 & 0.464 & 0.701 \\ \hline
shoe & 0.883 & 0.928 & 0.906 & \bf 0.931 & 0.827 & 0.896 \\ \hline
house & 0.904 & 0.926 & 0.854 & 0.902 & 0.897 & \bf 0.928 \\ \hline
scissors & 0.869 & \bf 0.903 & 0.842 & 0.886 & 0.873 & 0.872 \\ \hline
scrambledpix & 0.637 & 0.865 & 0.839 & \bf 0.914 & 0.754 & 0.874 \\ \hline
face & 0.833 & 0.801 & 0.820 & 0.818 & 0.835 & \bf 0.858 \\ \hline
bottle & 0.921 & 0.924 & 0.921 & \bf 0.948 & 0.787 & 0.943 \\ \hline
cat & 0.885 & 0.912 & 0.860 & \bf 0.915 & 0.783 & 0.908 \\ \hline
chair & 0.896 & 0.905 & 0.748 & \bf 0.922 & 0.837 & 0.867 \\ \hline
	 \end{tabular}
	 }
	 \vspace{7mm}
	 \centering
	 \caption{Dependence of correlation on MF method and visual stimulus $(K=90)$}
	 \label{tab:corr3}
	 \scalebox{1.0}[1.0]{
	 \begin{tabular}{|c|cccccc|}
	 \hline
IMG & PCA & FastICA & SparsePCA & SparsePCA & MOD & MOD \\
&     & & (low) & (high) & (low) & (high) \\ \hline \hline
rest (no image) & 0.679 & \bf 0.798 & 0.613 & 0.770 & 0.336 & 0.760 \\ \hline
shoe & 0.883 & 0.877 & 0.758 & 0.913 & 0.690 & \bf 0.921 \\ \hline
house & 0.904 & 0.881 & 0.848 & 0.918 & 0.682 & \bf 0.946 \\ \hline
scissors & 0.869 & 0.878 & 0.817 & 0.865 & 0.838 & \bf 0.915 \\ \hline
scrambledpix & 0.637 & 0.881 & 0.720 & 0.914 & 0.671 & \bf 0.932 \\ \hline
face & 0.769 & 0.863 & 0.760 & 0.855 & 0.708 & \bf 0.890 \\ \hline
bottle & 0.921 & 0.909 & 0.794 & 0.938 & 0.784 & \bf 0.944 \\ \hline
cat	& 0.885 & 0.910 & 0.828 & 0.909 & 0.840 & \bf 0.919 \\ \hline
chair & 0.896 & 0.882 & 0.756 & 0.890 & 0.618 & \bf 0.905 \\ \hline
	 \end{tabular}
	 }
\end{table}

\begin{figure}[htbp]
	 \begin{center}
      \includegraphics[scale=0.59]{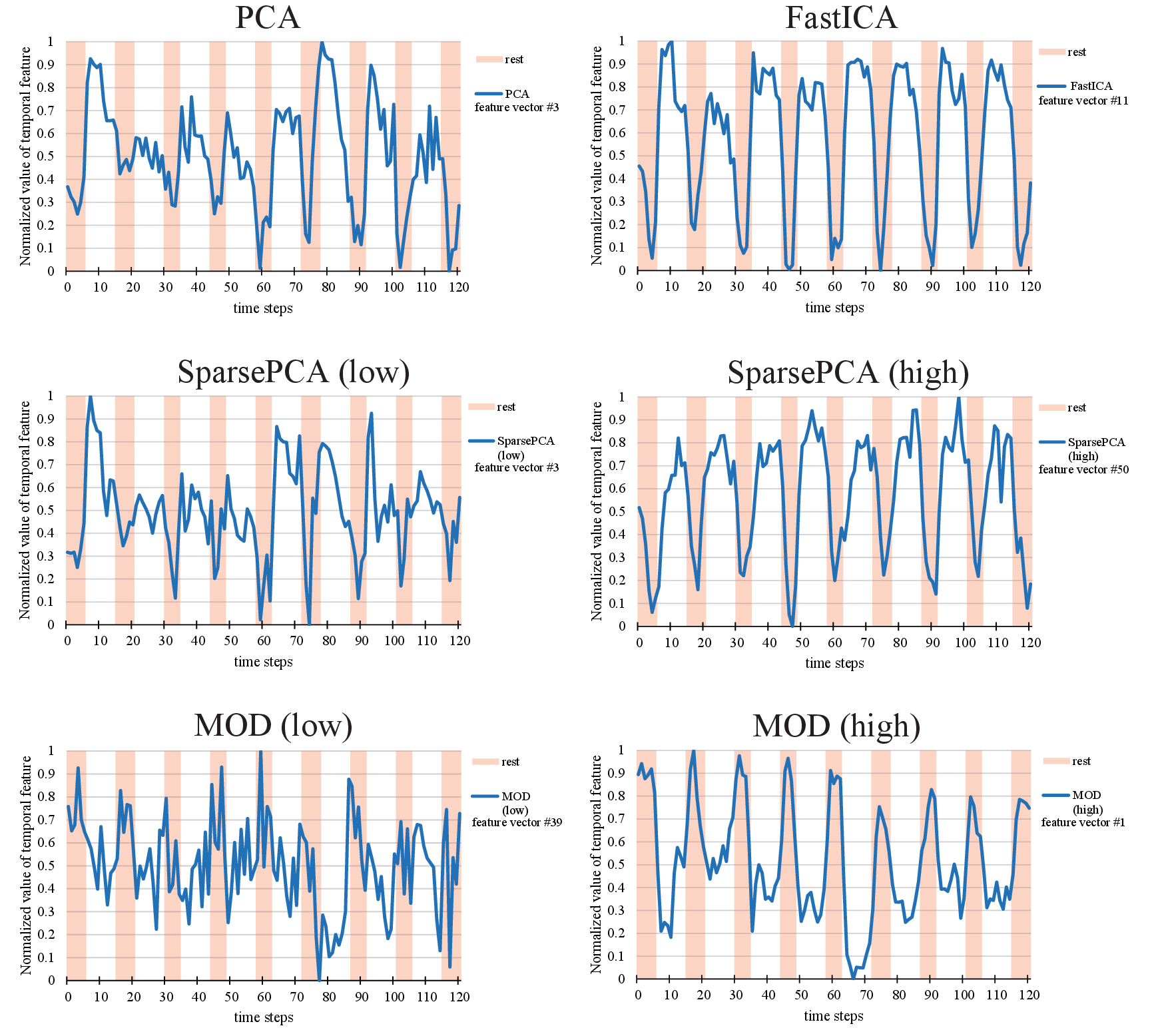}
	  \caption{The temporal column feature vectors $\bm d$ giving the largest value of correlation with
	  the resting state vector $\bm s^{\rm rest}$:
	  The vectors obtained by PCA, FastICA, SparsePCA, and MOD under $K=90$ are depicted.
	  The red area indicates the timing of resting state. The number for the feature vector means
	  the column number from the left in $\bm D$.}
	  \label{figure_correlation}
	 \end{center}
\end{figure}

To visualize the correlation, the temporal feature vectors giving the largest correlation with 
$\bm s^{\rm rest}$ under $K=90$ are depicted in Figure \ref{figure_correlation}.
In this figure, the temporal feature vectors are rescaled 
so that the maximum and minimum elements are 1 and 0, respectively.
From this figure, it is found that the temporal feature vectors by PCA, FastICA,
SparsePCA with high sparsity setting, and MOD with high sparsity setting are synchronized with the
timing of resting state in phase or anti-phase (=synchronization with some visual stimuli).
It is also confirmed that temporal feature vectors by sparse MF methods with high sparsity setting and FastICA 
are synchronized more accurately than PCA. 
These behaviors are consistent with the results of the correlations in the tables. 

\begin{figure}[htbp]
	 \begin{center}
      \includegraphics[scale=0.65]{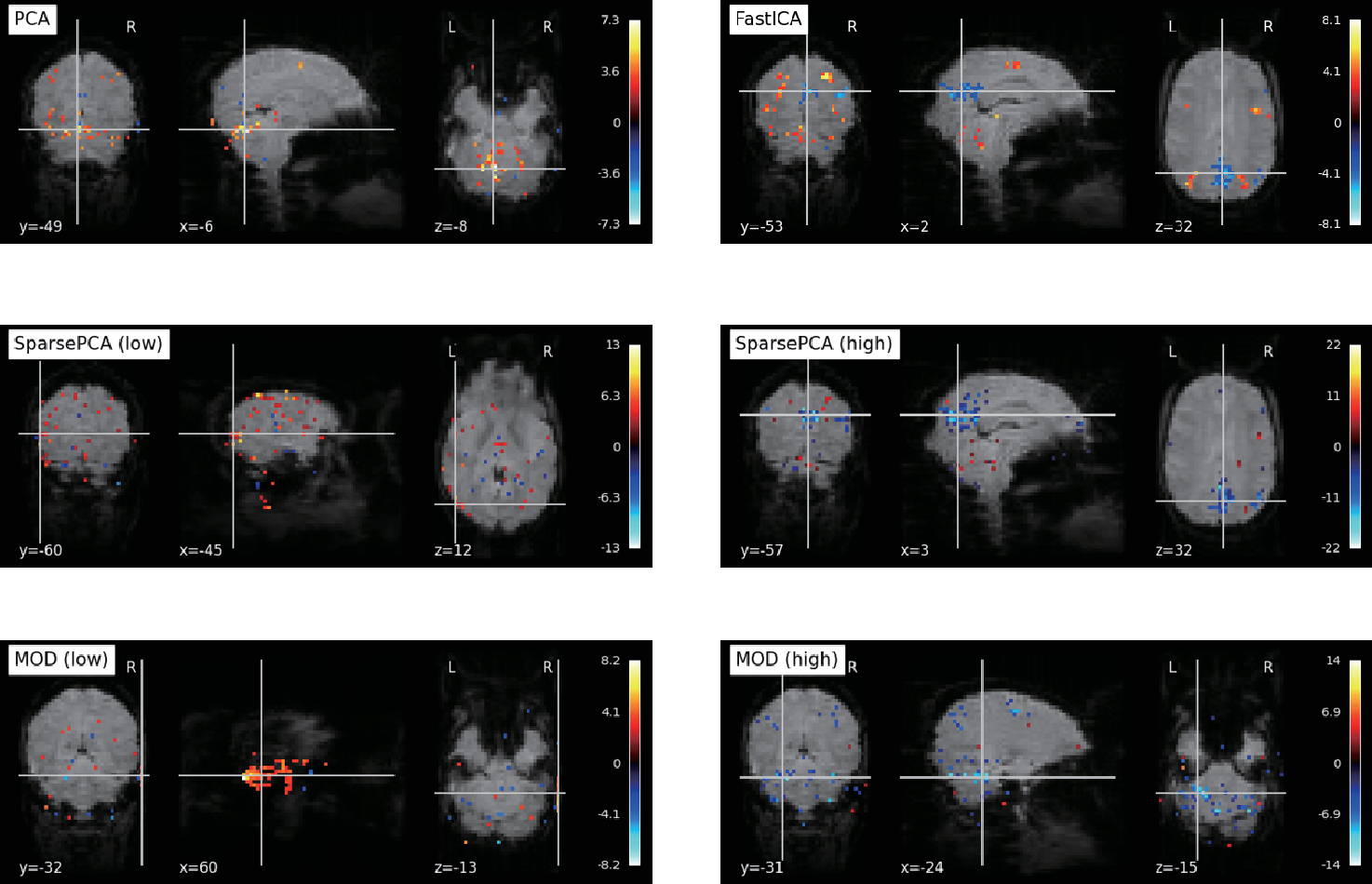}
	  \caption{The spatial maps (spatial feature vectors) visualized on the cross-sections of the brain: 
	  Each spatial map has one-to-one correspondence with the temporal feature vector with the largest correlation
	  (shown in Figure \ref{figure_correlation}).
	  The points in these figures 
	  are displayed according to the mapping between voxel position and each element in the spatial feature vector.
	  The color of the point indicates the value of the corresponding element.}
	  \label{spatial_map}
	 \end{center}
\end{figure}

In addition, the spatial maps (or equivalently spatial feature vectors) corresponding 
to these temporal feature vectors are illustrated in Figure \ref{spatial_map}.
For spatial map, the row vectors in $\bm X$ are rescaled to have zero mean and unit variance, 
and elements within 3 standard deviations of the mean are set to zero (called zero-filling hereafter).
For visualization, Nilearn (version 0.10.1) in Python library is used.
This figure will represent the active region under the resting state or the response to some visual stimuli.

Focusing on the results of PCA, FastICA, SparsePCA with high sparsity setting, and MOD with high sparsity setting, which show synchronization with the timing of resting state as aforementioned, activations are observed in the early visual cortex and near the cerebellum for FastICA or SparsePCA with high sparsity setting, and strong activations are observed in the cerebellum for PCA or MOD with high sparsity setting. 
Since these brain regions are known to be activated by visual stimuli, 
these results can be interpreted 
that specific MF methods can extract brain regions for information processing of visual stimuli.
By additional experiment, it is also confirmed that sparse MF methods with high sparsity setting and FastICA 
can identify the responding brain regions 
even without rescaling and zero-filling of $\bm X$, 
because the factorized matrix $\bm X$ can be obtained sparsely under these methods. 
In contrast, the factorized matrix $\bm X$ by PCA is not originally sparse. Accordingly,
strong activation near the cerebellum cannot be identified by PCA without rescaling and zero-filling.
This may be the reason why PCA yields smaller values of correlation than sparse MF methods with high sparsity setting
and FastICA.

%%%%%%%%%%
\subsection{Reconstruction error}
\label{sec4.3}

For validity of MF result, we evaluate reconstruction error of MF result.
Here we use the data from subject No. 2.

First, we compare the results by PCA and FastICA.
In Tables \ref{tab:ap20} and \ref{tab:ap50}, statistics of $\delta_{ij}$, Frobenius norm of $\bm \Delta$, 
and kurtosis of factorized matrix defined by equation (\ref{kurtosis}) 
from PCA and FastICA under $K=20, 50$ are shown. Kurtosis is measured for the concentration
of the matrix elements around zero. If kurtosis is large, the matrix is regarded as approximately sparse. 
\begin{equation}
\label{kurtosis}
{\rm kurtosis}(\bm X) = \frac{\frac{1}{KN} \sum_{i=1}^{K} \sum_{j=1}^{N} x_{ij}^4 }
{\left( \frac{1}{KN} \sum_{i=1}^{K} \sum_{j=1}^{N} x_{ij}^2 \right) ^2 } - 3.
\end{equation}
For FastICA, all statistics of $\delta_{ij}$ and Frobenius norm of $\bm \Delta$ under $K=20$ are larger than those under $K=50$. For PCA, these quantities excepting maximum of $\delta_{ij}$ are nearly independent of $K$.
By comparing the results between PCA and FastICA, FastICA slightly yields smaller Frobenius norm of $\bm \Delta$ and almost all statistics of $\delta_{ij}$. For kurtosis, FastICA gives larger value than PCA, which means that
factorized matrix by FastICA is approximately sparse.

Next, we compare the results by MOD, PCA, and FastICA.
In Tables \ref{tab:ap_mod20} and \ref{tab:ap_mod50}, statistics of $\delta_{ij}$, Frobenius norm of $\bm \Delta$, and sparsity of factorized matrix defined by equation (\ref{sparseness}) are shown for the result by MOD.
\begin{equation}
\label{sparseness}
{\rm sparsity}(\textbf{\textit{X}}) = \frac{\Sigma_{i=1}^{K} \Sigma_{j=1}^N \mathbb{I} [x_{ij} < \epsilon]}{KN},
\end{equation}
where $\epsilon$ is a small number and set to $10^{-5}$ in our study. Namely,
 {\rm sparsity}(\textbf{\textit{X}}) represents the fraction of almost-zero matrix elements (absolute value is smaller than $\epsilon$) in ${\bm X}$.  
From these tables, in both cases of $K=20, 50$, statistics of $\delta_{ij}$ and Frobenius norm of $\bm \Delta$ gradually increase when sparsity parameter $k_0$ decreases.
In both cases of $K=20$ (Tables \ref{tab:ap20} and \ref{tab:ap_mod20}) and $K=50$ (Tables \ref{tab:ap50} 
and \ref{tab:ap_mod50}),
MOD yields larger values than PCA or FastICA for all indices.
 
We also compare the results by SparsePCA and MOD. In Tables \ref{tab:as20} and \ref{tab:as50}, the results by SparsePCA under varying sparse parameter $\lambda$ are shown.
For comparison between SparsePCA and MOD, the value of $\lambda$ is chosen for which the resulting sparsity($\bm X$) is nearly equal to that under specific setting of $k_0$ in MOD.
In SparsePCA, sparser (or larger) setting of $\lambda$ does not necessarily mean larger indices of reconstruction error unlike MOD. 
In addition, from results by all MF methods, SparsePCA gives the largest reconstruction error among all MF methods.

Finally, we again discuss the property of the kurtosis by FastICA. By comparing the values of kurtosis 
by various MF methods, it is found that value of kurtosis by FastICA is comparable 
to the values by sparse MF methods. This indicates that approximately sparse factorized matrix is realized by FastICA. 

\begin{table}[htbp]
	 \centering
	 \caption{Reconstruction error of PCA and FastICA ($K=20$)}
	 \label{tab:ap20}
	 \scalebox{1.0}[1.0]{
	 \begin{tabular}{|l|cccccc|}
	 \hline
method & maximum & median & mean & variance & Frobenius norm & kurtosis \\ \hline \hline
PCA & 3.46 & 0.11 & 0.15 & 0.02 & $0.46 \times 10^3$ & 6.91 \\ \hline
FastICA & 3.36 & 0.10 & 0.15 & 0.02 & $0.45 \times 10^3$ & 18.69 \\ \hline
	 \end{tabular}
	 }
	 \vspace{3mm}
	 \centering
	 \caption{Reconstruction error of PCA and FastICA ($K=50$)}
	 \label{tab:ap50}
	 \scalebox{1.0}[1.0]{
	 \begin{tabular}{|l|cccccc|}
	 \hline
method & maximum & median & mean & variance & Frobenius norm & kurtosis \\ \hline \hline
PCA & 2.76 & 0.11 & 0.16 & 0.02 & $0.48 \times 10^3$ & 4.61 \\ \hline
FastICA & 1.87 & 0.08 & 0.11 & 0.01 & $0.34 \times 10^3$ & 17.71 \\ \hline
	 \end{tabular}
	 }
	 \vspace{3mm}
	 \centering
	 \caption{Reconstruction error of MOD ($K=20$)}
	 \label{tab:ap_mod20}
	 \scalebox{1.0}[1.0]{
	 \begin{tabular}{|r|ccccccc|}
	 \hline
$k_0$ & maximum & median & mean & variance & Frobenius norm & sparsity$(\bm X)$ & kurtosis \\ \hline \hline
15 & 4.12 & 0.64 & 0.67 & 0.17 & $1.73 \times 10^3$ & 0.25 & 5.06 \\ \hline
10 & 3.38 & 0.64 & 0.67 & 0.18 & $1.74 \times 10^3$ & 0.50 & 23.12 \\ \hline
5 & 4.13 & 0.65 & 0.68 & 0.18 & $1.77 \times 10^3$ & 0.75 & 17.72 \\ \hline
1 & 9.06 & 0.68 & 0.71 & 0.20 & $1.84 \times 10^3$ & 0.95 & 46.09 \\ \hline
	 \end{tabular}
	 }
	 \vspace{3mm}
	 \centering
	 \caption{Reconstruction error of MOD ($K=50$)}
	 \label{tab:ap_mod50}
	 \scalebox{1.0}[1.0]{
	 \begin{tabular}{|r|ccccccc|}
	 \hline
$k_0$  & maximum & median & mean & variance & Frobenius norm & sparsity$(\bm X)$ & kurtosis \\ \hline \hline
38 & 2.88 & 0.63 & 0.65 & 0.17 & $1.69 \times 10^3$ & 0.24 & 2.48 \\ \hline
25 & 2.99 & 0.63 & 0.66 & 0.17 & $1.70 \times 10^3$ & 0.50 & 8.61 \\ \hline
13 & 3.10 & 0.64 & 0.66 & 0.17 & $1.72 \times 10^3$ & 0.74 & 31.04 \\ \hline
1 & 5.41 & 0.67 & 0.70 & 0.19 & $1.82 \times 10^3$ & 0.98 & 147.80 \\ \hline
	 \end{tabular}
	 }
	\vspace{3mm}
	\centering
	\caption{Reconstruction error of SparsePCA ($K=20$)}
	\label{tab:as20}
	\scalebox{1.0}[1.0]{
	\begin{tabular}{|l|ccccccc|}
	 \hline
$\lambda$ & maximum & median & mean & variance & Frobenius norm & sparsity$(\bm X)$ & kurtosis \\ \hline \hline
0.1 & 8.83 & 0.77 & 0.82 & 0.28 & $2.14 \times 10^3$ & 0.27 & 18.21 \\ \hline
0.2 & 6.37 & 0.79 & 0.85 & 0.30 & $2.22 \times 10^3$ & 0.52 & 35.63 \\ \hline
0.4 & 6.76 & 0.79 & 0.84 & 0.30 & $2.20 \times 10^3$ & 0.75 & 61.65 \\ \hline
%1.0 & 10.37 & 0.80 & 0.86 & 0.32 & $2.27 \times 10^3$ & 17.33\\ %\hline
	\end{tabular}
    }
    \vspace{3mm}
	\centering
	\caption{Reconstruction error of SparsePCA ($K=50$)}
	\label{tab:as50}
	\scalebox{1.0}[1.0]{
	\begin{tabular}{|l|ccccccc|}
	 \hline
$\lambda$ & maximum & median & mean & variance & Frobenius norm & sparsity$(\bm X)$ & kurtosis \\ \hline \hline
0.07 & 5.76 & 0.77 & 0.82 & 0.27 & $2.12 \times 10^3$ & 0.27 & 7.81 \\ \hline
0.15 & 9.32 & 0.80 & 0.86 & 0.32 & $2.26 \times 10^3$ & 0.56 & 22.83 \\ \hline
0.25 & 7.77 & 0.79 & 0.85 & 0.31 & $2.24 \times 10^3$ & 0.77 & 61.49 \\ \hline
%1.0 & 8.21 & 0.80 & 0.86 & 0.32 & $2.23 \times 10^3$ & 0.93\\ \hline
	\end{tabular}
	}
\end{table}

%%%%%%%%%%
\subsection{Discussion}
\label{sec4.4}

We found that classification accuracy by PCA or sparse MF method (SparsePCA, MOD) 
with low sparsity setting tends to decrease with increasing $K$. 
On the other hand, sparse MF methods with high sparsity setting
and FastICA yield high accuracy regardless of the value of $K$ if an appropriate classifier is chosen.
In particular, sparse MF methods always give high accuracy regardless of the choice of the classifier.
The result of correlation also suggests that sparse MF methods 
with high sparsity setting and FastICA can extract temporal feature vector of a particular visual stimulus appropriately.
In contrast, MF without sparsity or sparse MF method with low sparsity setting 
does not necessarily give larger correlation with visual stimulus.
To summarize the result, we can conclude the strong possibility of {\it sparse}  information representation 
for visual stimulus in the whole human brain, because sparse MF methods and FastICA, which
also realizes approximate sparsity in factorized matrix, yield higher classification accuracy. 

For reconstruction error, FastICA gives the smallest values of statistical indices among all MF methods, and PCA gives the second smallest values in general.
Therefore, in feature extraction by MF, PCA and FastICA can keep as much information of the original matrix $\bm{Y}$ as possible.
The reason for the small reconstruction errors for PCA and FastICA can be attributed to the fact that they do not assume strict sparsity on the factorized matrix.
The sparse MF method requires reconstruction of the original matrix by matrix multiplication 
with limited number of nonzero elements, because many factorized 
matrix elements must be strictly zero. 
For FastICA, judging from large kurtosis,
the factorized matrix has many small elements close to zero, however these elements are not strictly zero.
Therefore, under PCA or FastICA, it is possible to use all elements in the factorized matrix to reconstruct the original matrix by matrix multiplication, and thus to obtain a solution with a small reconstruction error.

%%%%%%%%%%%%%%%%%%%%%
\section{Summary}
\label{sec5}

In this study, the validity of sparse cording in information processing for visual stimulus in the whole human brain has been 
investigated by the application of MF to task-related fMRI data and performance evaluation of various MF methods. 
We found that sparse MF methods (namely SparsePCA, MOD) with high sparsity setting and FastICA
realizing approximate sparsity yield high classification accuracy regardless of the classifier and large correlation.
We believe that this fact supports sparse coding hypothesis in information representation in the brain.

In future study, we will attempt to reconstruct other external stimuli using sparse MF. 
In addition, it is known that deep learning methods are often used for the estimation of 
synchronously active brain regions and the reconstruction of external stimuli,
for example variational autoencoder \cite{HAN2019} 
or deep MF \cite{QIAO2021,Wylie2021,Zhang2022a,Zhang2022b}.
However, the studies with deep learning did not focus specifically on sparsity.
Therefore, we will need to compare our results with sparse deep learning methods or non-sparse methods.
Finally, we will develop our findings into new feature extraction methods for the information processing in the brain.

%%%%%%%%%%
\section*{Acknowledgments}

We are grateful to the anonymous reviewer for the suggestion regarding the appropriate use of FastICA
in reviewing this article for publication.
This work is supported by KAKENHI Nos. 18K11175, 19K12178, 20H05774, 20H05776, and 23K10978.

%%%%%%%%%%
\bibliographystyle{unsrt}
\bibliography{article}

\end{document}